\documentclass[sn-basic,iicol]{sn-jnl}

\usepackage{amsmath,amssymb,amsthm}
\usepackage{graphics}
\usepackage{physics}
\usepackage{comment}
\usepackage{optidef}
\usepackage{graphicx}
\usepackage{booktabs}
\newcommand{\ZZ}{\mathbb Z}
\newcommand{\RR}{\mathbb R}

\usepackage{caption} 
\jyear{2022}%

\theoremstyle{thmstyleone}%
\theoremstyle{thmstyletwo}%

\theoremstyle{thmstylethree}%

\begin{document}

\title[Article Title]{Fixed interval scheduling problem with minimal idle time with an application to music arrangement problem}


\author*[1,2]{\fnm{Ludmila} \sur{Botelho}}\email{lbotelho@iitis.pl}

\author[2,3]{\fnm{\"Ozlem} \sur{Salehi}}\email{ozlemsalehi@gmail.com}

\affil*[1]{\orgdiv{Joint Doctoral School}, \orgname{Silesian University of Technology}, \orgaddress{\street{Akademicka 2a}, \city{Gliwice}, \postcode{44-100}, \country{Poland}}}

\affil[2]{\orgname{Institute of Theoretical and Applied Informatics, olish Academy of Sciences}, \orgaddress{\street{Bałtycka~5}, \city{Gliwice}, \postcode{44-100}, \country{Poland}}}

\affil[3]{\orgname{QWorld Association}, \city{Tallinn}  \country{Estonia}}


\abstract{The Operational Fixed Interval Scheduling Problem aims to find an assignment of jobs to machines that maximizes the total weight of the completed jobs. We introduce a new variant of the problem where we consider the additional goal of minimizing the idle time, the total duration during which the machines are idle. The problem is expressed using quadratic unconstrained binary optimization (QUBO) formulation, taking into account soft and hard constraints required to ensure that the number of jobs running at a time point is desirably equal to the number of machines. Our choice of QUBO representation is motivated by the increasing popularity of new computational architectures such as neuromorphic processors, coherent Ising machines, and quantum and quantum-inspired digital annealers for which QUBO is a natural input. An optimization problem that can be solved using the presented QUBO formulation is the music reduction problem, the process of reducing a given music piece for a smaller number of instruments. We use two music compositions to test the QUBO formulation and compare the performance of simulated, quantum, and hybrid annealing algorithms.}

\keywords{Interval scheduling, QUBO, annealing, music reduction}



\maketitle

\section{Introduction}\label{sec1}

Interval scheduling problems cover a broad range of problems arising in various areas, including production, scheduling, and logistics. One variant is the Operational Fixed Interval Scheduling Problem (OFISP), which is characterized as the problem of assigning a number of jobs, each with a fixed starting and ending time and weight, to a number of machines with the restrictions that each machine can handle a single job at a time and preemption is not allowed \citep{arkin1987scheduling,bouzina1996interval, KROON1995190}. The objective is to find an assignment of jobs to the machines with maximal total weight. The problem is studied in the literature under different names and different variants, including $k$-track assignment \citep{brucker1994thek} and $k$-coloring of intervals \citep{carlisle1995k}. We refer readers to \cite{kolen2007interval} and \cite{KOVALYOV2007331} for detailed surveys on the topic. Exact algorithms are known for solving OFISP in polynomial-time \citep{arkin1987scheduling, bouzina1996interval} based on solving a minimal cost flow problem.

In OFISP, a machine can be idle during a time period, which we call the idle-time. We introduce Operational Fixed Interval Scheduling Problem with Minimal-Idle Time (OFISP$_{\text {min-i}}$), where the goal is to find an assignment that maximizes the total weight while minimizing the total idle time. With this new condition, the optimal solution to the problem is no longer the same as the original problem. In this variant, we have a multi-objective problem where we would like to maximize the total weight of the completed jobs while we minimize idle time. The concept of minimal or no idle-time has been considered previously for flow shop problems in the literature \citep{ruiz2009scheduling, goncharov2009flow} but not for OFISP, up to our knowledge. A related problem is studied in \citep{chrobak2015scheduling,demaine2007scheduling}, where the authors consider minimizing the total number of gaps in the schedule with an assumption that all jobs have unit time.

This paper presents a quadratic unconstrained binary optimization (QUBO) formulation for the OFISP$_\text{min-i}$ problem. We express the problem by introducing binary variables to represent the selected jobs where the objective is to maximize the total weight of the selected jobs. Additional terms are introduced to ensure that the number of selected jobs at each time point is equal to the number of machines and that the idle time is minimized. A distinctive feature of our formulation from the existing ones is that the optimization procedure outputs the selected jobs, not the job-machine assignment. Hence after solving the QUBO, post-processing is performed to determine the job-machine assignment.

Our choice of QUBO, as opposed to other formulations like integer linear programming, is motivated by the fact that the incorporation of constraints through the penalty method naturally enables us to differentiate between soft and hard constraints. Furthermore, the number of novel architectures for solving QUBO models has been increasing over the years, such as quantum annealers \citep{johnson2011quantum}, neuromorphic processors \citep{corder2018solving}, coherent Ising machines realized by optical setup \citep{inagaki2016coherent} and quantum-inspired digital annealers \citep{tsukamoto2017accelerator}.

An optimization problem that can be targeted with OFISP$_{\text {min-i}}$ is the music reduction problem. The problem involves selecting self-contained parts of the song (phrases) played by individual instruments so that a music piece consisting of multiple instruments can be played with a smaller number of instruments of preference. The phrases should be selected in a way that overlapping phrases are not assigned to the same instrument. In addition, the phrases should reflect the characteristics of the song as much as possible, identified by their weights. Furthermore, since we want to preserve the flow and linearity of the song without unnecessary interruptions, this motivates us to select phrases leaving as minimum gaps as possible. The phrases played by instruments in a timeline can be considered analogically as jobs executed by machines in a timeline. Thus, the music reduction problem can be reduced to selecting jobs with minimal idle time to maximize the total weight.

We test the presented QUBO formulation on two classical music pieces: Suite No. 3 in D major, BWV 1068, Air, and Beethoven's Symphony No. 7 in A major, Op. 92, Second Movement. For solving the QUBO, we consider simulated annealing, quantum annealing, and hybrid quantum annealing. simulated annealing is a probabilistic search algorithm to approximate global optimum, used commonly with unconstrained problems defined over discrete search spaces \citep{kirkpatrick1983optimization}. Quantum annealing (QA) is a heuristic quantum algorithm with the same purpose relying on adiabatic quantum theorem \citep{apolloni1989quantum,kadowaki1998quantum, farhi2000quantum}. It can be performed by the devices provided by D-Wave company \citep{johnson2011quantum}, and problems expressed as QUBO formulations can be naturally targeted with QA. D-Wave also provides hybrid solvers that make use of classical optimization together with queries made to quantum annealers. We present the experimental results and compare the solution quality obtained from different solvers.

This paper is organized as follows: in Section \ref{sec2} covers the preliminary aspects of simulated and quantum annealing methods, the QUBO model, and reviews the integer linear program formulation for OFISP. In Section \ref{sec3} we present the QUBO formulation for OFISP and machines assignment after job selection. Following, on Section \ref{sec2}, we 4 the experiment for music reduction problem, with phrase identification, weight determination via information entropy, formulation as jobs, and results obtained for two different compositions. We conclude by Section by \ref{sec5} with a discussion on
future directions.

\section{Preeliminars}\label{sec2}
This section presents the basics of simulated and quantum annealing, QUBO formulation and the state-of-the-art integer linear program for the OFISP problem.

\subsection{Simulated and quantum annealing}

A valid approach to deal with the computational hardness of large combinatorial optimization problems is the \emph{simulated annealing} (SA) algorithm. The method was introduced by \cite{kirkpatrick1983optimization} to solve the Traveling Salesman Problem. SA works by emulating the process of annealing a solid by slowly lowering the temperature so that when eventually its structure is ``frozen'', this happens at a minimum energy configuration. This method can be regarded as a random walk on the search space. The algorithm starts at an initial state $x$ having cost $c = f(x)$ and then iterates to walk through the problem landscape. At each iteration $k$, it selects a random neighbor, which is accepted if it has a lower cost, becoming the current solution $c_{new}$. If the new solution has a higher cost, it is accepted with a probability determined by the temperature parameter $t_k$ and the difference between the existing and the new costs. The acceptance probability function is usually defined as:
\begin{equation}
P[Accept(t_k,\Delta)] = \text{min}\left( e^{- \Delta/t_k} \right)
\end{equation}
where $\Delta = c_{new}-c$. As the cooling process is carried out, $t_k$ is decremented, and the optimal solution is found with the help of thermal fluctuations. The SA is a metaheuristic method and can be viewed as an adaptation of Metropolis-Hastings algorithm. We refer readers to \cite{koulamas1994survey} for a more detailed survey on SA. 

Simulated annealing has been used to solve various problems including scheduling problems \citep{potts1991single,osman1989simulated,van1992job}, Travelling Salesman Problem and its variants \citep{vcerny1985thermodynamical,alfa19913}, graph coloring \cite{chams1987some}, and quadratic assignment problem \citep{burkard1983heuristic}.

\emph{Quantum annealing} \citep{apolloni1989quantum, kadowaki1998quantum} is a heuristic algorithm that runs in the framework of adiabatic quantum computing (AQC) \citep{farhi2000quantum}, targeting optimization problems. In AQC, a system starting in the lowest energy state (ground state) of some initial Hamiltonian $H_0$ (a mathematical operator that describes the system's energy) is likely to stay in the ground state throughout the evolution, given that the system is evolved slowly enough. Hence, if some problem Hamiltonian $H_P$ is introduced gradually to the system, it is likely that the system ends up in the ground state of $H_P$ at the end of the evolution time $T$. Mathematically, the evolution of the system is described by the time-dependent Hamiltonian
\begin{equation}\label{eq:evolution}
H(t)=\left(1-\frac{t}{T}\right) H_{0}+\frac{t}{T} H_{P}.
\end{equation}
 Note that quantum annealing is a physical process in an analog quantum device, as opposed to simulated annealing, and exploits quantum phenomena like tunneling and superposition.

The quantum annealers provided by D-Wave implement a problem Hamiltonian whose energy is expressed by an \emph{Ising model} of the form
\begin{equation}
    E(s) = \sum_{i} h_{i} s_{i}+\sum_{i<j} J_{i j} s_{i} s_{j},
\end{equation}
where $s$ is a spin configuration of variables $s_i \in \{-1,1\}$. Thus, one can use quantum annealing to solve optimization tasks expressed in terms of an Ising model or equivalently in the form of QUBO since the transformation between the two can be easily accomplished. Note that finding the minimum energy configuration of an Ising model is known to be NP-Hard. 

Quantum annealing has been used to solve optimization problems from different domains, including transportation \citep{salehi2022unconstrained, domino2021quadratic,neukart2017traffic}, automotive \citep{glos2022optimizing, yarkoni2021multi}, chemistry\citep{perdomo2012finding}, finance \citep{rebentrost2018quantum,venturelli2019reverse} and scheduling \citep{venturelli2015quantum, ikeda2019application, denkena2021quantum}. Recently, it has been also used in the scope of music theory for composing music \cite{chapter}. The speedup provided by QA for such problems is under a scientific debate and not evident \citep{boixo2014evidence,hauke2020perspectives,mandra2018deceptive}.

\subsection{Quadratic unconstrained binary optimization}
The quadratic unconstrained binary optimization (QUBO) is formally expressed by the optimization problem
\begin{equation}
    \min  x^T Q x,
\end{equation}
where $x$ is a vector of binary decision variables and $Q$ is a square matrix of real coefficients. By definition, the QUBO model has no constraints other than the requirement for the variables to be binary. However, many combinatorial optimization problems often include additional constraints that must be satisfied besides an objective function to be minimized. Many of these constrained models, such as integer linear programs or integer quadratic programs, can be effectively re-formulated as a QUBO model by introducing quadratic penalties into the objective function as an alternative to explicitly imposing constraints in the classical sense \citep{lucas2014ising,salehi2022unconstrained}. 

The significance of the ability of the QUBO model to encompass many problems in combinatorial optimization is enhanced by the fact that the QUBO model can be shown to be equivalent to the Ising model. The transformation between QUBO and Ising model can be performed easily using the mapping $x_{i} \leftrightarrow \frac{1-s_{i}}{2}$. QUBO formulations for many optimization problems are presented in \citep{lucas2014ising, glover2018tutorial}.  



\subsection{Integer linear program for OFISP}
In this section, we will review the existing integer linear program formulation for OFISP, which will give insights for the new formulation we present for OFISP$_\text{min-i}$.

Let us start by formally defining an integer linear program (ILP). In integer linear programming, the problems are formulated through some set of linear constraints over integer variables and a linear objective function to be minimized. An ILP problem is defined as 
\begin{alignat*}{3}
&\min \hspace{1.6em}&& \sum_{j} c_j y_j \\
&\text{subject to} \hspace{1em }&& \sum_{j} a_{ij}y_j \leq b_i, \hspace{1em} i=1,\dots,m \\
&{} &&y_j \geq 0, y_j \in \ZZ   
\end{alignat*}
where $a_{ij}\in \RR$, $b_j\in \RR$, $ c_j \in \RR $. Integer quadratic program (IQP) is defined analogously with a quadratic objective function and a set of linear constraints. Both ILP and IQP problems can be expressed as QUBO problems by first converting integer variables into binary variables and then using the penalty method, as mentioned previously.

Let us recall the ILP formulation for the Operational Fixed Interval Scheduling problem presented in  \cite{arkin1987scheduling, barcia2005k}.
The OFISP consists of assigning a set of jobs $J = \{b_1, \dots,b_N \}$ to a set of resources or machines $R =\{r_1,\dots,r_M \}$. Each job $b_i $ has a particular weight $w_i$, a fixed start time $s_i$, and an ending time $e_i$. We consider the time as discretized time units. 

Let $ G=(J,E) $ be the graph with vertices $ J $ such that edge $(i,j) \in E$ if jobs $ b_i $ and $ b_j $ are compatible, i.e. $[s_i,e_i]$ and $[s_j,e_j]$ do not intersect. Let $ x_{ij} $ be the binary variable such that
\begin{equation}
x_{ij} =   \begin{cases}%
1,      & \text{if job $b_i$ is assigned to machine $r_j$}\\
0, & \text{otherwise}
\end{cases}
\end{equation}
for $ i = 1, \dots ,N $ and $ j=1, \dots ,M  $. The ILP formulation is given as follows:
\begin{align}
\max\hspace{0.1in}&\sum_{i=1}^N \sum_{j=1}^M w_i x_{ij} \label{eq:ilpobjective}\\
\text{such that}~~&\sum_{j=1}^M x_{ij} \leq 1 \mbox{ for }i=1,\dots, N \label{eq:singlemachine}\\
&\sum_{i \in I} x_{ij} \leq 1 \mbox{ for }j=1,\dots,M
\end{align}
where $ I \subseteq J$ in any maximal independent set of $ G $.

\begin{figure*}
    \centering
    \includegraphics[width=0.9\textwidth]{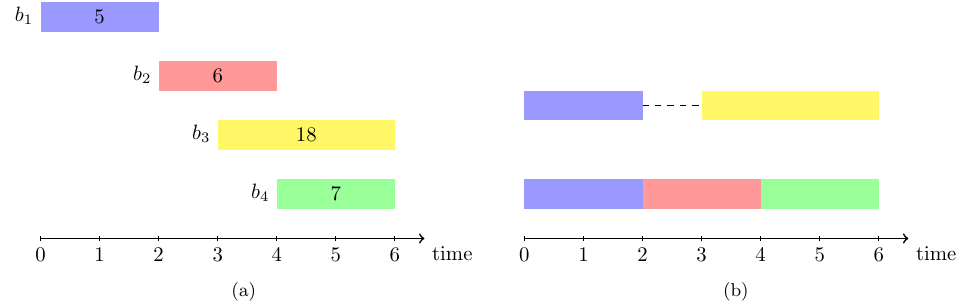}    

    \caption{On the left hand side, four jobs $b_1,b_2,b_3,b_4$ with the respective weights $5,6,18,7$ are displayed on the timeline. On the right-hand side, two different schedules are presented. The schedule at the top, with a total weight of 23, is the optimal solution to the OFISP problem, and there is a gap between the time points 2 and 3, indicating that the machine is idle during that time. The schedule at the bottom has no idle time, although the total weight is 18.}
    \label{fig:undesirable}
\end{figure*}

The first constraint ensures that each job is assigned to at most one machine. The second constraint ensures that incompatible jobs are not selected for the same machine. The drawback of this formulation is that it does not allow taking into consideration whether some machines are idle in a given time period. It can even be the case that all machines are idle at a time point. A case where this happens is depicted in Figure~\ref{fig:undesirable}. On the left-hand side, four jobs with their respective weights are depicted on the timeline. On the right-hand side, two different job assignments for a single machine are visualized. The optimal solution for the OFISP problem is displayed at the top, and one can observe the gap in the obtained solution. The solution in which the minimal time is minimized is given at the bottom.

\section{Optimization algorithm for OFISP$_{\text{min-i}}$}\label{sec3}

In this section, we first give the QUBO formulation for OFISP$_{\text{min-i}}$ problem and then discuss the classical postprocessing step.

\subsection{QUBO formulation for OFISP$_{\text{min-i}}$}\label{sec:qubo}

Let us define the binary variable $x_{i}$ such that
\begin{equation}
x_{i} =   \begin{cases}%
1,      & \text{if job $b_i$ is selected}\\
0, & \text{otherwise}
\end{cases}
\end{equation}
for $ i = 1, \dots ,N $.

Assuming that the time intervals for jobs are given by discrete time units (minutes, seconds), we need the following constraint:
 \begin{equation}\label{eq:hard_const}
            \sum_{i: ~ k\in[s_i,e_i]} x_i \leq M  \qquad \text{for $k =  0, \dots, K$. }
        \end{equation}
This constraint ensures that more than $M$ jobs are not selected for any time point and can be considered a hard constraint. Any solution violating this condition would be an infeasible solution.

We also would like to have exactly $M$ jobs assigned to each time point to minimize idle time, although this may not always be possible. Hence, we introduce the following constraint:
\begin{equation}\label{eq:soft_const}
    \sum_{i: ~ k\in[s_i,e_i]} x_i = M  \qquad \text{for $k =  0, \dots, K$.}
\end{equation}
Using the first constraint together with this additional condition which can be considered as a soft constraint, we penalize the cases where the number of selected jobs $x_i$ exceeds $M$ with a larger penalty than the case fewer jobs are selected. That also complies with the goal of minimization of idle time in case it is only possible to find a solution with idle time.

The goal is to maximize the total weight of the selected jobs:
\begin{equation}\label{eq:obj}
\sum_{i=1}^N w_ix_i.
\end{equation}
Incorporating the constraints presented in Eq.~\eqref{eq:soft_const} and Eq.~\eqref{eq:hard_const}, we present the following QUBO formulation:
\begin{align}
    & \sum_{i=1}^N w_ix_i + P_1 \sum_{k=1}^K  \left(\sum_{i: ~ k\in[s_i,e_i]} x_i + E(\xi_k) - M\right)^2 \nonumber \\
     &P_2 \sum_{k=1}^K  \left(\sum_{i: ~ k\in[s_i,e_i]} x_i - M\right)^2.\\
\end{align}

In the given formulation, $E(\xi_k)$ is the binary representation of the integer slack variable $\xi_k$, which is needed for converting the inequality constraint given in Eq.~\eqref{eq:hard_const} into equality. $P_1$ and $P_2$ are the penalty coefficients and $P_1$ should be selected larger than $P_2$. For the given formulation, the number of variables required is upper bounded by $O (N + K \log M)$: we need $N$ binary variables to represent $x_i$, and $K \log M$ variables to represent $K$ slack variables each ranging from $0$ to $M$. 

\subsection{Post-processing}

Note that the obtained solution does not give us information about the job machine assignment but only the list of selected jobs. Nevertheless, given the list of selected jobs, the assignment can be determined using a classical greedy algorithm typically used for \emph{interval partitioning}. The original problem aims to find an assignment with the minimum possible number of classrooms to schedule all the lectures with fixed starting and ending times. If we adapt this to our case, the lectures become jobs, and classrooms become machines. In this approach, the jobs are sorted according to their starting times and assigned greedily to machines. The algorithm has time complexity $O(N \log N)$ \cite{kleinberg2006algorithm}, where $N$ is the number of jobs. We describe the algorithm in detail in Algorithm \ref{alg:assignment} for completeness.

 \begin{algorithm} 
 	\caption{Pseudocode for job machine assignment}\label{alg:assignment}
 	\begin{algorithmic}[]
 		\Require $J$ -- the list of selected jobs
 		\State Sort $J$ by starting time in non-decreasing order
 		\State $m \gets 1$
 		\For{$j$ in $J$} 
 		\If{ $j$ is compatible with some machine $k \in \{1,\dots,m\}$}
 		\State Assign $j$ to $k$
 		\Else
 	    \State $m \gets m + 1$
 		\State Assign $j$ to $m$
            \EndIf
            \EndFor
 	\end{algorithmic}
 \end{algorithm}

\subsection{Enhancing the model} \label{sec:enhancing}

On top of the formulation presented above, one can introduce further constraints. For instance, it might be the case that we do not want particular jobs to be selected simultaneously. In such a case, it would be enough to include the term $P x_ix_j$ for the specific jobs $b_i$ and $b_j$, where $P$ is a suitable penalty coefficient. Note that one can prioritize some jobs by increasing their weights further. 

In the presented QUBO formulation, we assumed that all the machines were identical. The case of unidentical machines is commonly considered in the literature \citep{arkin1987scheduling}. In that case, we have the restriction about machines on which a job can be processed. Let us denote the set of machines on which job $b_i$ can be processed by $R_i \subseteq R$. To model the problem, we will introduce the binary variables indexed by the job number and the assigned machine, as in the case of the ILP for OFISP. Let $x_{ij}$ be the binary variable such that
\begin{equation}
x_{ij} =   \begin{cases}%
1,      & \text{if job $b_i$ is assigned to machine $r_j$}\\
0, & \text{otherwise.}
\end{cases}
\end{equation}
for $ i = 1, \dots ,N $ and $ j=1, \dots ,M  $. We need Eq.~\eqref{eq:singlemachine}, which forces each job to be assigned to a single machine, and the objective stays as in Eq.~\eqref{eq:ilpobjective}. Additionally, we need the following constraints:
\begin{equation}
    \sum_{i: ~ k\in[s_i,e_i]} \sum_{j=1}^M  x_{ij} \leq M  \qquad \text{for $k =  0, \dots, K$},
\end{equation}
\begin{equation}
    \sum_{i: ~ k\in[s_i,e_i]} \sum_{j=1}^M  x_{ij} = M  \qquad \text{for $k =  0, \dots, K$}.
\end{equation}

Now, to make sure that a job is not assigned to an incompatible machine, we add the terms $Px_{ij}$, for those $i,j$ pairs where $r_j \not \in  R_i$. Hence, any incompatible assignment is penalized. Note that the number of variables required for this formulation is $N\cdot M$ instead of $N$, without counting the slack variables.

\section{Music reduction}\label{sec4}
In this section, we describe the music reduction problem and discuss why it is a special case of the OFISP$_\text{min-i}$. To build the problem instance, the first step is the identification of the musical phrases and their weights. Subsequently, we present some experimental results.

Music reduction is the task of selecting particular parts - often called phrases - of a music piece involving multiple instruments to be played with a smaller number of instruments. A more complicated task is the music arrangement, where adjustments are done for the targeted instruments besides reduction. In previous research, several approaches to the automatic arrangement of ensemble or orchestral pieces for single instruments have been considered. A variety of methods for music reduction and phrase selection were applied previously such as machine learning \citep{chiu2009automatic}, state-transition models \citep{onuma2010piano}, hidden Markov model \citep{huang2012}, entropy optimization \citep{hori2014hmm}, and local boundary detection model \citep{you2019}. Besides the extraction of information, those works also consider the playability of the newly arranged pieces.

\subsection{Phrase identification}

A phrase in a melody is a sequence of notes that express a musical idea on its own. The ending of a phrase may be marked with relative lengthening of the last note, intensity or timbral change, or the presence of rests or pauses \citep{olsen2016constitutes,todd1985model,cambouropoulos2001local}. Identification of musical phrases has been considered an essential part of music perception by humans \citep{knosche2005perception,neuhaus2006effects}. Meanwhile, the automated detection of phrases has been considered by many researchers. Some approaches make use of machine learning techniques such as deep neural networks \citep{guan2018melodic} or use statistical modeling \citep{pearce2008comparison} and rule-based approaches which aim to identify the points of change in different musical parameters such as pitch and rhythm \citep{lerdahl1996generative, narmour1990analysis}. In \cite{you2019}, the phrase selection is based on computing the cost function with the information entropy of the phrase. 

In this paper, we use the Local Boundary Detection Model (LBDM) proposed by \cite{cambouropoulos2001local} for identifying musical phrases. First, sequences of pitch, interonset (duration between the beginning of two consecutive notes), and rest intervals are obtained from a given melodic sequence. For each interval, boundary strength values are calculated and combined to obtain the melody's overall local boundary strength profile. Boundary strength values are proportional to the degree of change between two consecutive intervals, and the boundary introduced on the larger interval is proportionally stronger. The peaks of this sequence indicate the local boundaries.  

To identify the peaks of the sequence, one needs to set a threshold value. However, based on the threshold value, the phrases' length may become too small or too large. To overcome this issue, we developed the algorithm described in Algorithm \ref{alg:phrases}, which performs a search for the threshold value. The algorithm and the detailed explanation can be found in Appendix~\ref{sec: phrasealg}.

\subsection{Determining the weight for each phrase}\label{sec:entropy}

The goal of music reduction is to highlight a song's most distinctive features and capture the main melody while reducing the number of instruments. The main melody usually consists of pitch and rhythm rich in information. More formally, such melodic phrases have higher entropy regarding information theory. In order to quantify the amount of information in a phrase, we used the approach presented in \cite{you2019}, which identifies pitch and interonset interval (IOI) entropy. The pitch entropy corresponds to the frequency variety in a scale occurring in a sequence of notes comprising a phrase. In the case of chords, which contain multiple pitches, we only take into account the highest note in the chord for entropy calculation because listeners tend to focus on the notes with the highest pitches.

Mathematically, the entropy for a random variable $X$ with possible values ${x_1, \dots, x_n}$ is given by
\begin{equation}
    H(X)=-\sum_{i=1}^{n} P\left(x_{i}\right) \log _{2} P\left(x_{i}\right),
\end{equation}
where $P(x_i)$ is the probability that $X$ takes value $x_i$. Suppose that we represent with $x_i$ the possible pitch values that can occur in a phrase. Then, the probabilities $P\left(x_{i}\right)$ are computed by
\begin{equation}
    P\left(x_{i}\right)=\frac{n_{i}}{N},
\end{equation}
where $n_{i}$ is the number of occurrences of the $i$th pitch in the phrase and the phrase contains $N$ notes. The melody entropy for each phrase is calculated based on this probability calculation.

In addition to pitch, rhythm is also a factor in determining the amount of information a phrase contains. Compared with the duration of notes, the IOI is easier to perceive. Similarly to the pitch, we calculate the IOI entropy based on the its probability, with $x_i$ as the possible IOI value in a set of $N$ different IOIs with $n_i$ occurrences in a phrase.

The underlying assumption is that compared to the accompaniment, the rhythm in the main melody is more complicated and more difficult to predict. That means the combined pitch and IOI entropy of the phrase in the main melody part should be the largest.

\subsection{Solving the music reduction problem}
\begin{figure*}[h]
    \centering
    \includegraphics[width=0.8\textwidth]{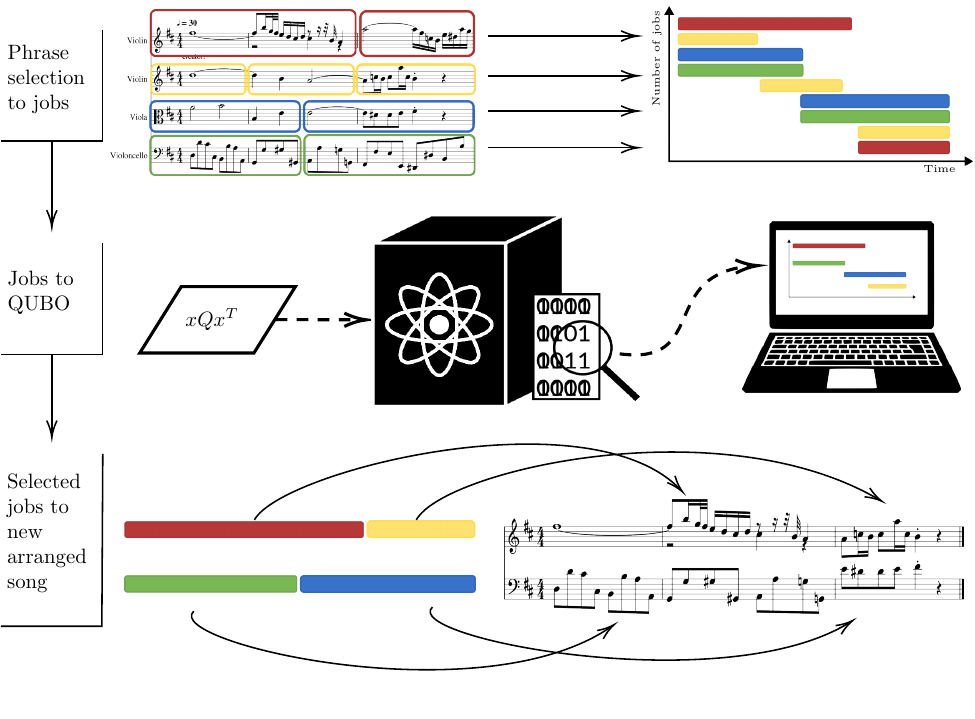}           
    \caption{The flowchart for solving music reduction problem as an application of OFISP$_{\text {min-i}}$.}
    \label{fig:workflow}
\end{figure*}

Let us summarize the overall procedure for solving the music reduction problem. Suppose that the original song contains $ N $ tracks, and the goal is to reduce the number of tracks to $ M $. We first identify the phrases using Algorithm \ref{alg:phrases}. Next, we determine the weight of each phrase as explained in Sec.~\ref{sec:entropy}. The list of phrases is the list of jobs in the OFISP$_{\text{min-i}}$, where the starting end ending times are the start and end measures of the phrase. The number of machines is simply $M$, the number of tracks we want to obtain after the reduction. The weight of each job is the information entropy of the corresponding phrase. Once the problem instance is created, then we create the QUBO formulation based on Sec.~\ref{sec:qubo} and use an annealing algorithm to solve the problem. After obtaining the list of selected jobs, we use the post-processing given in Algorithm \ref{alg:assignment} to obtain the assignment of jobs to machines. Equivalently, we get the phrase track assignment. The procedure is summarized in Fig.~\ref{fig:workflow}.

\subsection{Results}\label{sec:results}
We chose two classical compositions for our experiments: Suite No. 3 in D major, BWV 1068, Air, and Beethoven's Symphony No. 7 in A major, Op. 92, Second Movement. The music sheets are encoded as MIDI files with public domain licenses and parsed with \texttt{music21} \citep{music21} library. Our goal was to reduce the given music piece to two tracks. To start with, we classically processed the phrase selection algorithm, identified the phrases, and created the QUBO formulation using \texttt{pyqubo} \citep{pyqubo} library.

We conducted the annealing experiments with the available solvers provided by D-Wave, which included simulated, quantum, and hybrid annealers. The parameters for simulated annealing used in our experiment are the number of reads $n_r = 1000$ and the number of sweeps $n_s = 1000$. We used two quantum processing units (QPUs) in the quantum annealing experiments. The first one, \texttt{Advantage\_system4.1}, provides 5627 qubits interconnected by Pegasus graph topology. The second device, \texttt{Advantage2\_prototype1.1}, has 563 qubits with the underlying Zephyr topology. The tested parameters are discussed further below. For the hybrid solver, no parameters are set, and it returns a single solution.

When running a problem on D-Wave QPUs, the variables should be mapped to the QPU architecture, as the underlying graph representing the interactions in the QPU is not fully connected. This process is known as minor embedding, where multiple physical qubits represent a single logical qubit named a chain. The coupling between those qubits is called the chain strength, which should be accordingly set so that it is not too large to override the actual problem, yet it is not small so that the chain is not broken \citep{willsch2022benchmarking}. The problems were automatically embedded via D-Wave's native probabilistic embedding algorithm. 

For the quantum annealing solvers, we vary the parameters for chain strength as $c_r = 0.1,0.2,0.3$ and the annealing time as $t = 100,500,1000,2000$. The number of reads and the annealing time are subject to the relation $n_r \cdot t <10^6$, and we picked the maximum possible number of reads based on this relation.

The solutions in which hard constraints defined in Eq. \ref{eq:hard_const} are violated are considered infeasible, and the violations of Eq.\ref{eq:soft_const} are deemed as soft violations. The results point to two solutions; the first is based on the sample with the highest entropy, and the second is based on the sample with the minimum number of soft violations. We decided to proceed with higher entropy solutions to maximize the information from the original song. The code and its description used to generate the experiment is available on https://doi.org/10.5281/zenodo.7410349.

\paragraph{Suite No. 3 in D major, BWV 1068, Air}
For this composition, which comprehends four tracks with a length of 19 measures, we identified the set of jobs from the 41 identified phrases. From this, the number of QUBO variables is 80 

\begin{figure*}[h!]
    \centering
    \includegraphics[width=0.8\textwidth]{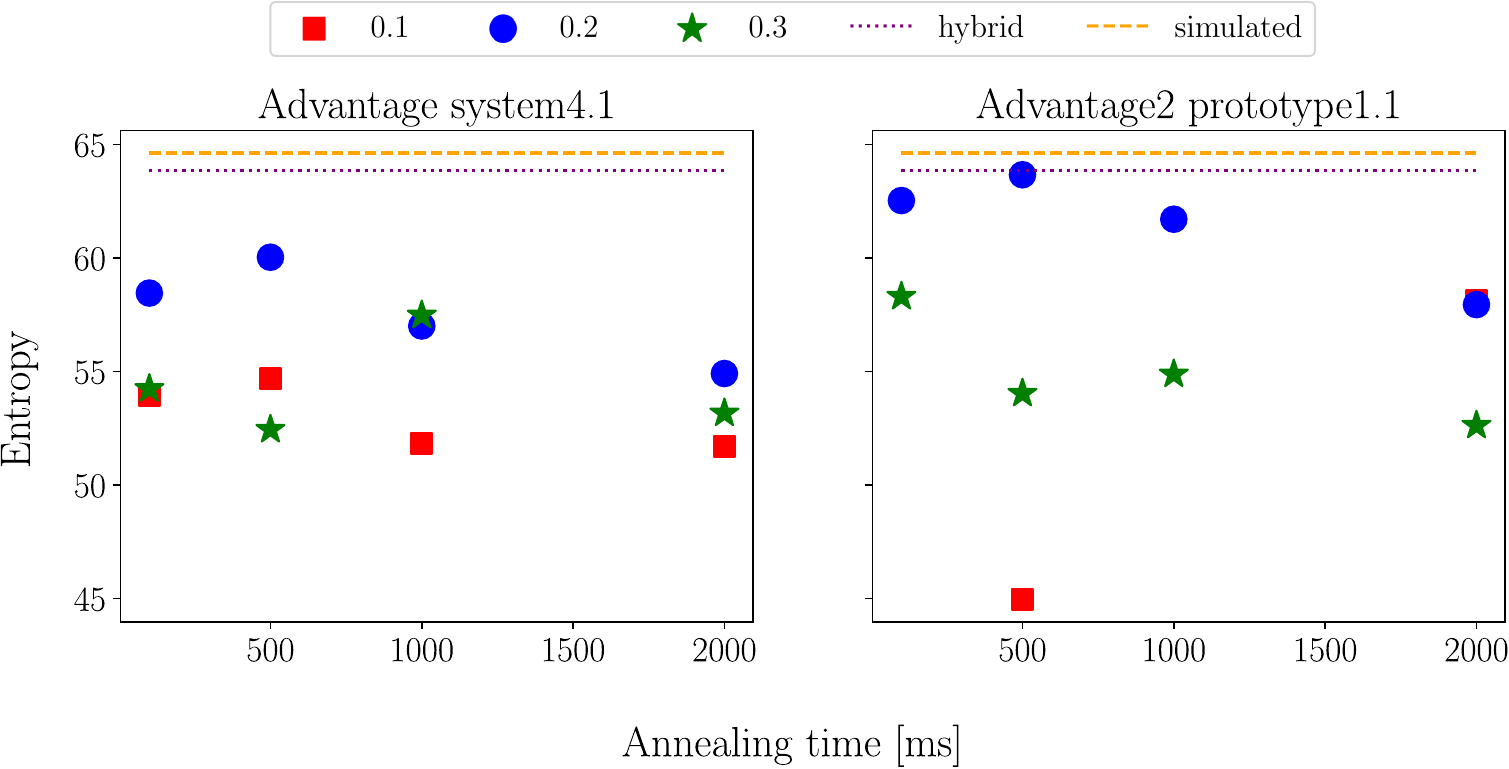}
    \caption{Experiment results for entropy varying chain strength and annealing times. The dashed lines correspond to the simulated annealing best result and the dotted lines to the hybrid solver solution. The square, round, and star dots correspond to chain strength values $0.1$, $0.2$, and $0.3$, respectively.}
    \label{fig:bach_plot}
\end{figure*}

The results are presented in Fig.~\ref{fig:bach_plot}. We compared the QPU results with simulated annealing and hybrid solver, which returned feasible solutions with no soft constraint violation and higher entropy. Feasible solutions were obtained for both \texttt{Advantage\_system4.1} and \texttt{Advantage2\_prototype1.1} QPUs, with soft violations. The number of physical variables for \texttt{Advantage\_system4.1} varied in a range of 107 to 130 qubits and for \texttt{Advantage2\_prototype1.1} from 106 to 126. The performances of both QPUs for chain strength $c_r = 0.2$ were absent of soft constraint violations and yielded the highest entropy. The number of soft and hard constraint violations can be found in Tables~\ref{tab:bach1} and \ref{tab:bach2} in Appendix~\ref{sec:A2}.


\paragraph{Symphony No. 7 in A major, Op. 92, Second Movement}
We decided to use for the second experiment a more challenging composition with 12 tracks of length 276 measures. We identified 591 phrases to build the set of jobs and, respectively, 1056 QUBO variables. 

The results for simulated annealing and Hybrid Solver returned feasible solutions with entropy values of $463.05$ and $441.99$, respectively. The solutions contain around the same amount of soft constraint violations, around $16\%$ of the measures. Due to the number qubit requirements, the problem did not fit into \texttt{Advantage2\_prototype1.1}. The number of physical variables required to solve the problem lay in the range of 2700 to 3020 for \texttt{Advantage system4.1}; however, no feasible solutions were obtained. For more details about the number of violated constraints, we refer to Tables~\ref{tab:beethoven1} and \ref{tab:beethoven2} in Appendix~\ref{sec:A2}.







\section{Conclusions and outlook}\label{sec5}

In this paper, we introduced a new scheduling problem variant called the Operational Fixed Interval Scheduling Problem with Minimal Idle Time, where the goal is to assign jobs to machines to minimize idle time while maximizing the total weight of the completed jobs. We presented a QUBO formulation for the problem and demonstrated an application for the music reduction problem. We provided some experimental results using quantum and hybrid solvers of D-Wave and simulated annealing.

The experiments conducted by the simulated and hybrid annealers indicate that the QUBO formulation proposed for the model is valid. In addition, the number of physical variables needed to encode the QUBO to the QPU compared to the number of logical variables suggests that the problem is suitable for solving in quantum devices due to having a sparse $Q$ matrix representation. Nevertheless, the quantum annealing experiments did not provide promising results, in particular for the larger problem. Note that the \texttt{Advantage2\_prototype1.1} is currently a prototype under development with a new hardware topology, and it seems to work better on the smaller problem instance compared to D-Wave Advantage. 

The hybrid solver's performance was comparable to simulated annealing, noting that simulated annealer returned higher entropy solutions. Considering the undergoing effort by D-Wave on hybrid solvers \citep{hybrid}, they can be alternative tools for solving large problem instances in the future. Furthermore, the presented model is suitable to be used by quantum-inspired annealers designed for solving combinatorial optimization problems, such as Fujitsu digital annealers \citep{tsukamoto2017accelerator}. 

Regarding the music reduction problem, observations in the conducted experiments suggest that one can further tune the model. Taking into account the enhancements proposed in Sec.~\ref{sec:enhancing}, one can avoid particular phrases being played simultaneously, which might create dissonance, or give higher weights to phrases from specific instruments. Furthermore, for a given phrase, one can indicate which instruments can play the phrase at the cost of having a larger number of variables.

\backmatter

\bmhead{Acknowledgments}

LB and \"O.S. acknowledge support from National Science Center under grant agreement 2019/33/B/ST6/02011.

The project was initiated under the QIntern program organized by QWorld, therefore we would like to thank the organizers of the program.

We would like to thank Adam Glos and Jaroslaw Adam Miszczak for their valuable comments.

\begin{appendices}

\section{Phrase selection algorithm}\label{sec: phrasealg}

 \begin{algorithm*}
 	\caption{Pseudocode for phrase identification}\label{alg:phrases}
 	\begin{algorithmic}
 		\Require $bs$ -- the list of boundary strength values, $k$ -- maximum number of measures in a phrase 
 		\State $min_{bs}\gets \min\{bs\}$, $max_{bs} \gets \max\{bs\}$
 		\State $t \gets (min_{bs} + max_{bs}) /2$
 		\State $found_t \gets None$ 
 		\While{true}{}
 		\State $flag  \gets false$
 		\State  $peaks \gets find\_peaks(bs,t)$
 		\If{ $peaks=[~]$}
 		\State $max_{bs} \gets t$
 		\State $t \gets (min_{bs} + max_{bs}) /2$
 		\State \textbf{continue}
 		\EndIf
 		\State $phrases \gets find\_phrases(peaks)$
 		\If {$length(p) > k$ for some $p~in~phrases $}
 		\State $max_{bs} \gets t$
 		\State $t \gets (min_{bs} + max_{bs}) /2$
 		\If {$t-min_{bs}< \epsilon$}
 		\If {$found_t \neq  None$}
 		\State  $peaks \gets find\_peaks(bs,found_t)$
 		\State $phrases \gets find\_phrases(peaks)$
 		\Return phrases
 		\Else
 	    \State $min_{bs}\gets \min\{bs\}$, $max_{bs} \gets \max\{bs\}$
 		\State $t \gets (min_{bs} + max_{bs}) /2$
 		\State $k \gets k+1$
 		\EndIf
 		\EndIf
 		\Else
 		\State $found_t \gets t$
 		\State $min_{bs} \gets t$
 		\State $t \gets (min_{bs} + max_{bs}) /2$
 		\If {$max_{bs} -t < \epsilon$}
 		\Return phrases
 		\EndIf
 		\EndIf
 		\EndWhile
 	\end{algorithmic}
 \end{algorithm*}

The algorithm takes into account the number of maximum measures allowed in a phrase identified by the user. The initial threshold value is determined as the average value in the list, and upper and lower bounds are determined as the minimum and maximum values in the list. Based on this threshold value, the peaks are identified. If no peaks can be identified for the selected threshold value, then the threshold value is decreased, and the upper bound is updated. In the next step, peaks are used to generate the phrases. Note that the peaks correspond to individual notes, but for simplicity, we assume that the phrase ends at the end of the measure containing the note. If all phrases have a length smaller than the allowed number of measures, then we record this threshold value (as it is working), and the threshold is increased, and the lower bound is updated. If the threshold value is too close to the upper bound, then we return the found phrases. If there is a phrase longer than the allowed number of measures, then the threshold value is decreased, and we go back to the stage where we identify the peaks. If the threshold value is decreased so that it is close to the lower bound, then this implies that the segmentation can not result in the desired number of allowed measures. If there is a previously recorded threshold, then the phrases are returned based on that threshold. Otherwise, the number of allowed measures in a phrase is incremented by 1, and the threshold, upper and lower bounds are reinitialized. 

At the end of the procedure, we end up with a list of phrases. Each phrase is identified by the measure it starts and ends. If any phrase contains only silences, then those are removed.

\bigskip
\section{Tables}\label{sec:A2}

In this section, we present the tables containing information on the number of violated constraints for the experiments discussed in Sec.~\ref{sec:results}. The number of each type of constraint is equal to the number of measures on the composition. Therefore, for  Suite No. 3 in D major, BWV 1068 there are 19 hard constraints and  19 soft constraints. For Symphony No. 7 in A major, Op. 92, Second
Movement, 276 hard constraints, and 276 soft constraints. 
\begin{table}[ht]
\begin{center}
\begin{tabular}{cccc}
\toprule
\multicolumn{4}{c}{\texttt{Advantage\_system4.1}}                                                       \\ \midrule
\multicolumn{1}{c}{}
& \multicolumn{3}{c}{Chain Strength}                       \\ \midrule
\multicolumn{1}{c}{Annealing Time} & \multicolumn{1}{c}{0.1} & \multicolumn{1}{c}{0.2} & 0.3 \\ \midrule
\multicolumn{1}{c}{100} 
& \multicolumn{1}{c}{0}   & \multicolumn{1}{c}{0}   & 0   \\ \midrule
\multicolumn{1}{c}{500}            & \multicolumn{1}{c}{4}   & \multicolumn{1}{c}{0}   & 1   \\ \midrule
\multicolumn{1}{c}{1000}           & \multicolumn{1}{c}{3}   & \multicolumn{1}{c}{0}   & 1   \\ \midrule
\multicolumn{1}{c}{2000}           & \multicolumn{1}{c}{1}   & \multicolumn{1}{c}{0}   & 1   \\ \bottomrule
\end{tabular}
\end{center}
\caption{Number of soft constraint violations for Suite No. 3 in D major, BWV 1068, \texttt{Advantage\_system4.1}.}
\label{tab:bach1}
\end{table}

\begin{table}[ht]
\begin{center}
\begin{tabular}{cccc}
\toprule
\multicolumn{4}{c}{\texttt{Advantage2\_prototype1.1}}                                                   \\ \midrule
\multicolumn{1}{c}{}               & \multicolumn{3}{c}{Chain Strength}                       \\ \midrule
\multicolumn{1}{c}{Annealing Time} & \multicolumn{1}{c}{0.1} & \multicolumn{1}{c}{0.2} & 0.3 \\ \midrule
\multicolumn{1}{c}{100}            & \multicolumn{1}{c}{4}   & \multicolumn{1}{c}{0}   & 0   \\ \midrule
\multicolumn{1}{c}{500}            & \multicolumn{1}{c}{2}   & \multicolumn{1}{c}{0}   & 0   \\ \midrule
\multicolumn{1}{c}{1000}           & \multicolumn{1}{c}{5}   & \multicolumn{1}{c}{0}   & 0   \\ \midrule
\multicolumn{1}{c}{2000}           & \multicolumn{1}{c}{0}   & \multicolumn{1}{c}{0}   & 1   \\ 
\bottomrule
\end{tabular}
\end{center}
\caption{Number of soft constraint violations for Suite No. 3 in D major, BWV 1068, \texttt{Advantage2\_prototype1.1}.}
\label{tab:bach2}
\end{table}

\begin{table}[ht]
\begin{center}
\begin{tabular}{cccc}
\toprule
\multicolumn{4}{c}{Soft Violations}                                                             \\ \midrule
\multicolumn{1}{c}{}               & \multicolumn{3}{c}{Chain Strength}                       \\ \midrule
\multicolumn{1}{c}{Annealing Time} & \multicolumn{1}{c}{0.1} & \multicolumn{1}{c}{0.2} & 0.3 \\ \midrule
\multicolumn{1}{c}{100}            & \multicolumn{1}{c}{149} & \multicolumn{1}{c}{157} & 163 \\ \midrule
\multicolumn{1}{c}{500}            & \multicolumn{1}{c}{153} & \multicolumn{1}{c}{162} & 167 \\ \midrule
\multicolumn{1}{c}{1000}           & \multicolumn{1}{c}{151} & \multicolumn{1}{c}{161} & 161 \\ \midrule
\multicolumn{1}{c}{2000}           & \multicolumn{1}{c}{149} & \multicolumn{1}{c}{162} & 170 \\ \bottomrule
\end{tabular}  
\end{center}
\caption{Number of soft constraint violations for for Symphony No. 7 in A major, Op. 92, Second Movement, \texttt{Advantage\_system4.1}.}
\label{tab:beethoven1}
\end{table}

\begin{table}[ht]
\begin{center}
\begin{tabular}{cccc}
\toprule
\multicolumn{4}{c}{Hard Violations}                                                            \\ \midrule
\multicolumn{1}{c}{}               & \multicolumn{3}{c}{Chain Strength}                       \\ \midrule
\multicolumn{1}{c}{Annealing Time} & \multicolumn{1}{c}{0.1} & \multicolumn{1}{c}{0.2} & 0.3 \\ \midrule
\multicolumn{1}{c}{100}            & \multicolumn{1}{c}{26}  & \multicolumn{1}{c}{40}  & 33  \\ \midrule
\multicolumn{1}{c}{500}            & \multicolumn{1}{c}{25}  & \multicolumn{1}{c}{45}  & 40  \\ \midrule
\multicolumn{1}{c}{1000}           & \multicolumn{1}{c}{34}  & \multicolumn{1}{c}{42}  & 54  \\ \midrule
\multicolumn{1}{c}{2000}           & \multicolumn{1}{c}{26}  & \multicolumn{1}{c}{41}  & 54  \\ \bottomrule
\end{tabular}    
\end{center}
\caption{Number of hard constraint violations for for Symphony No. 7 in A major, Op. 92, Second Movement, \texttt{Advantage\_system4.1}.}
\label{tab:beethoven2}
\end{table}




\end{appendices}


\bibliography{sn-bibliography}


\end{document}